\documentclass[12pt]{article}
\usepackage{amsmath}
\usepackage{amsfonts}

\newcommand{\bean}{\begin{eqnarray*}}
\newcommand{\eean}{\end{eqnarray*}}
\newcommand{\ed}{\end{document}}

\newcommand{\be}{\begin{equation}}
\newcommand{\ee}{\end{equation}}
\newcommand{\barr}{\begin{array}}
\newcommand{\earr}{\end{array}}
\newcommand{\bea}{\begin{eqnarray}}
\newcommand{\eea}{\end{eqnarray}}

\newcommand{\proof}{\medskip\noindent{\it Proof.}\quad }
\newcommand{\qed}{\hfill \fbox{}\medskip}

\begin{document}
\title{Deformation quantization of
Poisson manifolds in the derivative expansion.}
\author{A.V.Bratchikov
 \\ Kuban
State Technological University,\\ 2 Moskovskaya Street, Krasnodar,
  350072, Russia
} \date{} \maketitle

\begin{abstract}
Deformation quantization of
Poisson manifolds
is studied
within the framework
of an expansion in powers of derivatives of  Poisson structures.
We construct the Lie group associated with a Poisson  bracket algebra
which defines 
a second order deformation
in the derivative expansion.
\end{abstract}

PACS 2.40 Gh


\section  {Introduction}
Let $X$ be a phase space with the phase variables $x^i,\,i=1,...,N,$
and the Poisson bracket 
\bea \label{U}
\{x^i,x^j
\}=
\omega^{ij}(x).
\eea
Let $\tilde A=A[[t]]$
be the space of formal power series in
a variable $t$
with coefficients in $A=C^\infty(X).$
A star product
is a $R[[t]]$ linear associative product on
$\tilde A,$  defined for $f,g\in A$ by
\bean \label {or}
f\ast g=gf+\sum_{k=1}^\infty t^k
c_k(f,g),
\eean where $c_k(f,g)$ are bidifferential operators.

The star product is a basic object of the deformation quantization
\cite {BFFLS}.
A formula for
the star product  on an
arbitrary Poisson manifold was found by Kontsevich \cite {Kon}.
For symplectic Poisson manifolds one can also use Fedosov's
construction of deformation quantization \cite {F}.

The star product for linear Poisson structures
\cite {G}  can be derived from
the Baker-Campbell-Hausdorff (BCH) formula \cite {Kat}. It is different
from Kontsevich's product being a part of the last one.

In the present paper an  expansion  of the star product
in powers of the derivatives of Poisson structure is studied.
We find the explicit expression which defines a second order deformation in the derivative
expansion.This expression is given by an infinite series in variable $t$ as well as in the
number of first order derivatives of $\omega^{ij}$.
It is based on the BCH formula for the Poisson bracket algebra (\ref {U}).

The paper is organized as follows.
In the next section we introduce notations and rewrite  the
associativity equation in the form which is convenient for our
purposes. In Sec.3 we define an infinite-dimensional Lie group
associated with the Poisson bracket algebra (\ref {U}).
In Sec. 4 we show that the
corresponding BCH series defines a second order deformation
quantization in the derivative expansion.

\section {Derivative expansion}

For any $f,g \in A$ the star product can be computed by the
equation
\bea   \label {stm}
f(x)\ast g(x)=
f(\frac {\partial} {\partial \alpha})g
( \frac {\partial} {\partial \beta})
e^{Q(\alpha \cdot x ,\beta \cdot x)} |_{\alpha=\beta=0},
\eea
where $\alpha \cdot x=\alpha_ix^i,\alpha_i\in R,$ and
\bean   \label {stmi}
e^{Q(\alpha \cdot x,\beta \cdot x)}=e^{\alpha \cdot
x}\ast e^{\beta \cdot x}.
\eean


The  associativity equation for the star product reads
\bea
\label {assoy}
(f\ast g)\ast h=
f\ast (g\ast h)
\eea
for all
$f,g,h\in \tilde A.$
For $e^{\alpha \cdot x},e^{\beta \cdot x}$ and
$e^{\gamma \cdot x}$ this equation
can be written in the form
\bea \label {a}
e^{Q(\alpha \cdot x,\beta \cdot x)}|_{x=\frac {\partial} {\partial
\pi}} e^{Q(\pi \cdot x,\gamma \cdot x)}|_{\pi=0}=
e^{Q(\beta \cdot x,\gamma \cdot x)}|_{x=\frac
{\partial} {\partial \pi}} e^{Q(\alpha \cdot x,\pi \cdot x)}|_{\pi=0}.  \eea

It is easy to see that equations
(\ref {assoy}) and
(\ref {a})  are equivalent:
\begin{multline*}
(f\ast g)\ast h 
= f(\frac
{\partial}
{\partial \alpha})
g (\frac
{\partial} {\partial \beta})
h(\frac
{\partial}
{\partial \gamma})
e^
{Q(\alpha \cdot x,\beta \cdot x)}
|_{x=\frac {\partial} {\partial \pi}}
e^{Q(\pi \cdot x,\gamma \cdot x)}|_{\pi=\alpha=\beta=\gamma=0}= \\
=
f(\frac {\partial}
{\partial \alpha})
g (\frac
{\partial} {\partial \beta})
h(\frac
{\partial}
{\partial \gamma})
e^{Q(\beta \cdot x,\gamma \cdot
x)}|_{x=\frac {\partial} {\partial \pi}} e^
{Q(\alpha \cdot x,\pi \cdot
x)}|_{\pi=\alpha=\beta=\gamma=0}=
f\ast (g\ast h).
\end{multline*}

Let $B$ be the algebra with respect to the pointwise
multiplication generated by
$\omega$ and its derivatives.
Let
$B_n, n\in N,$ be the ideal of
$B$ generated by $n-$th and higher derivatives of
$\omega$ and
$\tilde B_n =B/B_n.$
When associativity equation (\ref {a}) holds on
$\tilde B_n$ we say that
product (\ref {stm}) defines an
$n-$th order deformation of $A$ in the derivative expansion.
If $\psi-\varphi\in B_n$ we write $\psi=\varphi+
O(\partial^n \omega).$

We are interested in the expansion of
$Q
(\alpha \cdot x,\beta \cdot x)\in B$ in
powers of derivatives of the Poisson structure
$\omega=(
\omega^{ij})$
\bean \label {assqi}
Q
(\alpha \cdot x,\beta \cdot x)
=
(\alpha +\beta) \cdot x+
\sum_{n=1}^\infty
{Q_{n}
(\alpha \cdot x,\beta \cdot x)},
\eean
where $Q_n\in B_{n-1},  Q_n\notin B_{n}.$
One checks that
\bean
Q_1(\alpha \cdot x,\beta \cdot x)=
\frac {t} {2} \alpha_i\omega^{ij}(x)\beta_j
\eean
defines a first order deformation in the derivative expansion.




\section  {Lie groups associated with Poisson bracket algebras}
Let $V^n=V\otimes V \otimes \ldots \otimes V$ ($n$ copies),$V^0=R,$
be the space over $R$ which is generated by 
$$x^{i_1}\otimes  x^{i_2}\otimes \ldots \otimes x^{i_n}. 
$$
Let $K$ denote the tensor algebra
$$
K=\bigoplus_{n=0}^\infty V^n.
$$
One can make $K$ into a Lie algebra $K$  by taking as Lie product $[a,b]=a\otimes b- b\otimes a.$ 
The Lie elements
\bean
[x^{i_n},\ldots ,[x^{i_2},x^{i_1}]]
\eean
generate a Lie subalgebra $
L$ of $
K.$  

Let $H (.\,,.)$  denote the BCH series in $
L$
\bean
\label {h}
H (a, b)=
a+b+
\frac {1}
{2} [a,b]+ \frac {1} {12} [a,[a,b]]+ \frac {1}
{12}[b,[b,a]]+\ldots .
\eean
The function $H(.\,,.)$ satisfies the equations (see e.g. \cite {B}) 
\bean  \label {sort}
H(a,0)= H(0,a)=a,\qquad H(a,-a)= 0,
\eean
\bea  \label {sorti}
H(H(a,b),c)= H(a,H(b,c))\qquad  \mbox {for all}\,\,\, a,b,c\in L.   
\eea

Let $T: L \to \tilde A$ be the linear function  which is defined by 
\bean
T([x^{i_n},\ldots ,[x^{i_2},x^{i_1}]])= t^{n-1}
\{x^{i_n},\ldots ,\{x^{i_2},x^{i_1}\}\}
\eean
where
$$\{f(x),g(x)\}= \partial_i f(x)
\omega^{ij}(x) \partial_j g(x)$$ 
and $\partial_i=  \partial /{\partial x^i}.$
The following lemma is easily proved by induction.

\vspace{3mm}
\noindent

LEMMA {\it The map  $T$ is a homomorphism of Lie algebras.
}
\vspace{3mm}
\noindent

The homomorphic images  of equations (\ref {sorti}) in $\tilde A$ 
reads
\bean  \label {sort}
H'(a',0)= H'(0,a')=a',\qquad H'(a',-a')= 0,
\eean
\bea  \label {som}
H'(H'(a',b'),c')= H'(a',H'(b',c'))
\eea
where $a'=T(a), b'=T(b), c'=T(c)$ and $H'(.\,,.)$ is the BCH series for the Poisson bracket $t\{.\,,.\}$
\bean
\label {h}
H'(f, g)=f+g+
\frac {t}
{2} \{f,g\}+ \frac {t^2} {12} \{f,\{f,g\}\}+ \frac {t^2}
{12}\{g,\{g,f\}\}+\ldots.
\eean

Equations (\ref {som}) enables us to define the Lie group $G$ which is generated by the functions $e^{\alpha\cdot x}$
and the product 
$$
e^{\alpha\cdot x}\circ e^{\beta\cdot x}= e^{H'({\alpha\cdot x},{\beta \cdot x})}
$$
Each element of $G$ can be written in the form
\bea \label {rno}
{\prod_{i=1}^s}
(e^{\gamma^{(i)}\cdot x}\circ)
=e^{\sum_{i=1}^s \gamma^{(i)}\cdot x}(1+O(t))
\eea
for some values of $s$ and group parameters $
(\gamma^{(1)},\gamma^{(2)},\ldots,\gamma^{(s)}).$
Here
\bean
\prod_{i=1}^s
(e^{\gamma^{(i)}\cdot x}\circ)=
e^{\gamma^{(1)}\cdot x} \circ
e^{\gamma^{(2)}\cdot x}\circ \ldots \circ
e^{\gamma^{(s)}\cdot x} .
\eean

It is easy to see by making use of (\ref {rno}) that 
$$G^t=\{u\in G |u=1+O(t)\}$$ is a normal subgroup of $G$
and $G/G^t$ is isomorphic to the
translation group $\hat R^N=\{\chi_\alpha | \chi_\alpha =
e^{\alpha\cdot x} \}.$ Therefore $G$  can be
treated as a deformation of $\hat R^N.$

\section  {A second order deformation
in the derivative expansion
}
Let us define the $R[[t]]$ linear product
on $\tilde A$
\bea
\label {sr}
f(x)\diamond g(x)=
f(\frac {\partial} {\partial \alpha})g
( \frac {\partial} {\partial \beta})
e^{H'(\alpha \cdot x,\beta \cdot x)}|_{\alpha=\beta=0}.
\eea
It can be also written in the form
\bean  \label {sram} f(x)\diamond
g(x)= e^{ M(\frac {\partial} {\partial \xi}\cdot x, \frac {\partial}
{\partial \eta} \cdot x)} f(\xi) g(\eta)|_{\xi=\eta=x},
\eean
where
\bean
M(f,g)= H'(f,g)-f-g
\eean

For the Poisson bracket with constant $\omega^{ij}$
$$M(\frac {\partial} {\partial \xi} \cdot x,
\frac {\partial}
{\partial \eta} \cdot x) =
\frac {t}
{2} \omega^{ij} \frac {\partial} {\partial \xi^i}
\frac {\partial} {\partial
\eta^j}.$$
In this case (\ref {sr}) reproduces the Groenewold-Moyal
product
\bean  \label {star6}
f(x)\diamond g(x)=
e^{\frac {t}
{2} \omega^{ij} \frac {\partial} {\partial \xi^i}
\frac {\partial} {\partial
\eta^j}} f(\xi) g(\eta)|_{\xi=\eta=x}.
\eean
Product
(\ref {sr})
coincides also with the star product  for linear Poisson structures
\cite {Kat}.

Expanding (\ref {sr}) in powers of $t$ one gets
\begin{multline*}
f\diamond g=fg+ \frac {t} {2}
\omega^{ij}\partial_i f\partial_j g+ \frac {t^2} {8} \omega^{ij}
\omega^{kl}\partial_i \partial_k f \partial_j \partial_l g+\\ +\frac
{t^2} {12} \omega^{ij}\partial_j\omega^{kl}(\partial_i \partial_k f
\partial_l g-
\partial_k f \partial_i\partial_l g)+
O(t^3).
\end{multline*}
This means
that (\ref {sr}) defines a second order deformation in
the variable $t$ \cite {Kon,PV}.


\vspace{3mm}
\noindent

THEOREM
{\it Formula (\ref {sr}) gives a second order deformation of $A$ in
the derivative expansion.}

\proof
Let $l(x)$ be a Lie series in $x^i.$ It can be written as 
\bean
l(x)=\alpha\cdot x +\sum_{n=2}^\infty
\alpha_{i_1i_2\ldots i_n}\{x^{i_n},\ldots ,\{x^{i_2},x^{i_1}\}\}
\eean
for some $\alpha_{i_1i_2\ldots i_n}\in R.$
One checks that \bea \label {v} l(x+a)=l(x)+ \partial_i
l(x) a^i+O(\partial^2 \omega) \eea and for any function $f(x)$ \bea
\label {d} \{l(x),f(x)\}=\{{\partial_i} l(x)x^i,f(x)\} + O(\partial^2
\omega).  \eea From this it follows
\begin {multline} \label {anu} H'(l(x),\gamma \cdot x)= 
l(x)+\gamma \cdot x + M(\partial_i l(x)x^i,\gamma \cdot x)+ O(\partial^2 \omega).
\end {multline}

Let us compute the left-hand side of equation
(\ref {a}) for $
Q(\alpha \cdot x,\beta \cdot x)=
H' (\alpha \cdot x,\beta \cdot x).$
Using (\ref {v}) and (\ref {anu})  one gets
\begin {multline}
\label {ony}
e^{H'(\alpha \cdot x,\beta \cdot x)}|_{x=\frac {\partial} {\partial
\pi}} e^{H'(\pi \cdot x,\gamma \cdot x)}|_{\pi=0}=
e^{H'(\alpha \cdot x,\beta \cdot x)}|_{x=x+\frac {\partial} {\partial
\pi}} e^{\gamma \cdot x+M(\pi \cdot x,\gamma \cdot x)}
|_{\pi=0}=\\
=e^{\gamma \cdot x+ H'(\alpha \cdot x,\beta \cdot x)+
M(\partial_i H'(\alpha \cdot x,\beta \cdot x)x^i,\gamma \cdot x)}+
O(\partial^2 \omega)=
e^{H'(H'(\alpha \cdot x,\beta \cdot x),\gamma \cdot x))}+O(\partial^2
\omega).
\end {multline}

By similar computations of the right-hand side of (\ref
{a}) we obtain
\bea
\label {olly}
e^{H'(\beta \cdot x,\gamma \cdot x)}|_{x=\frac {\partial} {\partial
\pi}} e^{H'(\alpha \cdot x,\pi \cdot x)}|_{\pi=0}=
e^{H'(\alpha \cdot x,H'(\beta \cdot
x,\gamma \cdot x))}+O(\partial^2 \omega).
\eea
From (\ref {ony}),(\ref {olly}) and
(\ref {som}) it follows that $H'(\alpha \cdot x,\beta \cdot x)$
satisfies equation (\ref
{a})  modulo  $
O(\partial^2
\omega).$
This means that
\bean
\label {assoyv}
(f\diamond g)\diamond h=
f\diamond (g\diamond h)+
O(\partial^2
\omega)\eean
for all
$f,g,h\in \tilde A.$
\qed


\section  {Conclusion
}

In the present article we have constructed the Lie group associated
with a Poisson bracket algebra.
We have found that the
corresponding BCH series defines a second order deformation of
Poisson manifolds in the derivative expansion.

It would be interesting to find a third order deformation. 
It defines a deformation of quadratic Poisson structures.

\bigskip


\end{document}